\documentclass[twocolumn,preprintnumbers,amsmath,amssymb]{revtex4}
\usepackage{graphicx}
\usepackage{dcolumn}
\usepackage{bm}
\usepackage{epsfig}
 {
 {
 \def\f{\frac}
 \def\dis{\displaystyle}
 \def\leqq{\leqslant}

 \def\etaj{\eta_j^{~}}
 \def\mB{\overline{m}}
 
 \def\psiB{\overline{\psi}}
 \def\phiT{\widetilde{\phi}}

\begin{document}

\title{Neutrino Mass and Baryon Asymmetry from Dirac Seesaw}

\author{Pei-Hong Gu\,$^{1,2}$}

\email{pgu@ictp.it}

\author{Hong-Jian He\,$^{1}$}

\email{hjhe@tsinghua.edu.cn}

\affiliation{
 $^{1}$Center for High Energy Physics, Tsinghua
       University, Beijing 100084, China\\
 $^{2}$The Abdus Salam International Centre for Theoretical Physics,
       Strada Costiera 11, I-34014 Trieste, Italy}

 \begin{abstract}
 We extend the visible content of the standard model (SM) with a hidden
 sector composed of three right-handed singlet neutrinos and one singlet Higgs.
 These extra singlets are charged under a new $U(1)_{X}^{~}$ gauge symmetry
 while the SM particles are not. Two heavy scalar doublets are introduced to
 play the role of the messengers between the visible and hidden sectors. The
 neutrinos naturally acquire tiny Dirac masses because
 the ratio of weak scale over the heavy messenger masses
 is highly suppressed. Furthermore, the heavy messengers
 simultaneously generate baryon asymmetry of the universe through their
 out-of-equilibrium {\tt CP}-violating decays.
 \\[2mm]
 \hfill [\,{\tt hep-ph/0610275}\,]
  \end{abstract}

\maketitle

 The tiny neutrino masses and the matter-antimatter asymmetry
 in the universe pose two major challenges to particle physics and cosmology.
 This indicates the necessity of supplementing to the existing theory
 certain new ingredients, which have been hidden from the direct
 experimental observations so far\,\cite{Frank}.

 In this paper, we propose a novel \textit{Dirac Seesaw} model
 to obtain the tiny neutrino masses and
 generate the observed baryon asymmetry by extending the visible
 content of the standard model (SM) with a hidden sector
 which is prescribed by a $U(1)_{X}^{~}$ gauge symmetry and
 composed of three right-handed singlet neutrinos plus one singlet Higgs.
 Two heavy scalar doublets are introduced to play the role of
 messengers between the visible and hidden sectors.

 The field content of the proposed $SU(2)_{L}^{}\otimes
 U(1)_{Y}^{}\otimes U(1)_{X}^{}$ model is defined in Table\,\ref{charge},
 where $\psi_{L}^{}$, $\phi$, $\nu_{R}^{}$, $\chi$, $\eta$ are the
 left-handed lepton doublets, the Higgs doublet, the right-handed
 singlet neutrinos, the singlet Higgs and the heavy scalar doublets,
 respectively. All other unlisted SM fields are $U(1)_{X}^{}$ singlets.
 In this Table we have also suppressed the generation indices for simplicity.
 It is clear that the conventional
 Yukawa couplings of the left-handed lepton doublets and the
 right-handed singlet neutrinos with the light Higgs doublet $\phi$ are forbidden
 because the right-handed neutrinos have $U(1)_{X}^{}$
 charge while all the SM particles do not. Namely, the
 sector composed of the right-handed neutrinos and the Higgs singlet
 is hidden from the visible SM sector. However, the heavy
 scalar doublets $\eta$, which join not only the
 SM gauge group $SU(2)_L\otimes U(1)_{Y}^{}$ but
 also the new $U(1)_{X}^{}$, can bridge the two sectors.
 In this sense, we may regard the heavy scalar doublets as ``messengers".

\begin{table}
\begin{center}
\begin{tabular}{c|ccc}
 \hline\hline
 &&& \\[-2mm]
  ~~~Fields          \quad\quad & ~~\,$SU(2)_{L}^{}$ \quad\quad & $   ~~U(1)_{Y}^{}
  \quad\quad $ & $   ~~U(1)_{X}^{}~~~   $
  \\[1.5mm]
  \hline
 &&& \\[-2mm]
  ~~~$\psi_{L}^{}$  \quad\quad &     ~~\,{\bf 2}    \quad\quad & $  -\frac{1}{2}
  \quad\quad $ & $        ~\,0        $~~~
  \\[1.5mm]
  ~~~$\phi$         \quad\quad &     ~~\,{\bf 2}    \quad\quad & $  -\frac{1}{2}
  \quad\quad $ & $        ~\,0        $~~~
  \\[1.5mm]
  ~~~$\nu_{R}^{}$   \quad\quad &     ~~\,{\bf 1}    \quad\quad & $      ~\,0
  \quad\quad $ & $  -\frac{1}{2}   $~~~
  \\[1.5mm]
  ~~~$\chi$         \quad\quad &     ~~\,{\bf 1}    \quad\quad & $      ~\,0
  \quad\quad $ & $  +\frac{1}{2}   $~~~
  \\[1.5mm]
  ~~~$\eta$         \quad\quad &     ~~\,{\bf 2}    \quad\quad & $  -\frac{1}{2}
  \quad\quad $ & $  +\frac{1}{2}   $~~~
  \\[-2mm]
  &&&
  \\ \hline \hline
 \end{tabular}
 \caption{The field content related to the generation of the Dirac
 neutrino masses. All other SM fields are $U(1)_X$ singlets
 and the family indices have been omitted for simplicity.}
 \label{charge}
 \end{center}
 \end{table}

 \begin{figure}
 \vspace{19.1cm}
 \psfig{file=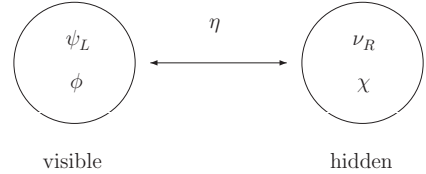, bbllx=4cm, bblly=-22.1cm, bburx=18cm,
 bbury=-8.1cm, width=8cm, height=8cm, angle=0, clip=0}
 \vspace{-23.5cm}
 \caption{\label{messenger}
 The heavy scalar doublets $\eta$ act as the
 messengers between the visible and hidden sectors.
 The other SM fields and the family indices have been omitted for
 simplicity. }
 \end{figure}

 \begin{figure}
 \vspace{4.5cm}
 \epsfig{file=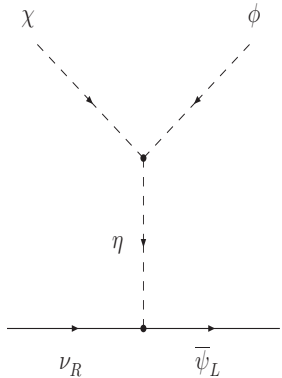, bbllx=4cm, bblly=6.0cm, bburx=18cm,
 bbury=20.0cm, width=8cm, height=10cm, angle=0, clip=0}
 \vspace{-9.0cm}
 \caption{\label{massgeneration} The generation of
 Dirac neutrino masses. Here the family indices have been
 suppressed. }
 \end{figure}

 We can thus write down the relevant interaction Lagrangian for
 generating the Dirac neutrino masses. It involves the messenger
 scalars $\eta$, the SM-like doublet $\phi$ and the singlet scalar
 $\chi$\,,
 \begin{eqnarray}
 \label{yt}
 \mathcal{L}& \supset &
 -y\,\psiB_L^{}\eta
 \nu_{R}^{}+
 \rho \chi\eta^{\dagger}_{}\phi
 +{\rm h.c.}\,,
 \end{eqnarray}
 where $\,y\,$ is the Yukawa coupling. The cubic scalar coupling
 $\rho$ has mass dimension-1 and is set to be real
 after a proper phase rotation.
 From Fig.\,\ref{massgeneration}, we note that at low energy we can
 integrate out the heavy messengers $\eta$ and obtain the
 following effective dimension-5 operator,
 \begin{eqnarray}
 \label{dim-5D}
 \dis
 {\cal L}^{(5)}_{\rm Dirac} ~=\,
 -y\f{\rho}{M_{\eta}^2}\psiB_L^{}\phi\chi\nu_R^{~}
 \,+\, {\rm h.c.}
 \,,
 \end{eqnarray}
 where $M_{\eta}^{}$ denotes the heavy messenger masses.
 So, once the Higgs
 fields $\phi$ and $\chi$ get nonzero vacuum expectation values (VEVs)
 after the symmetry breaking, the neutrinos naturally acquire small Dirac
 masses,
 \begin{eqnarray}
 \label{diracmass}
  M_{\nu}^{}
  ~\simeq~ y\frac{\,\rho\langle \phi \rangle
  \langle \chi \rangle\,}{M_{\eta}^{2}}
  ~=~ y\frac{\,\rho vu\,}{M_{\eta}^{2}}\,.
 \end{eqnarray}
 Here
 ~$\langle\phi\rangle\equiv v$~ and ~$\langle\chi\rangle\equiv u$~
 represent the scalar VEVs.
 From minimizing the scalar potential in Appendix-A,
 we have also shown that
 the heavy messengers will acquire tiny VEVs,
 ~$\left<\eta\right> \simeq
   \rho uv /M_{\eta}^{2} $\,,\,
 after $SU(2)_{L}^{}\otimes
 U(1)_{Y}^{}\otimes U(1)_{X}^{}$ is broken to $U(1)_{\rm em}^{}$.
 This confirms Eq.\,(\ref{diracmass}) due to the mass formula
 \,$M_{\nu} = y\left<\eta\right>$\, from Eq.\,(\ref{yt}).

 The vacuum expectation value
 $\,\langle\chi\rangle =u$\, breaks $U(1)_X$ and
 is expected to be nearby the weak scale,
 so we will set ~$u = \mathcal{O}(\textrm{TeV})$\,.\,
 As for the cubic scalar coupling $\rho$,
 we consider a reasonable case where
 $\rho $ is of order $M_{\eta}^{}$ or
 less. Under this setup, it is easy to see from Eq.\,(\ref{diracmass})
 that the Dirac neutrino masses will be efficiently
 suppressed by the ratio of the weak scale over the heavy
 messenger masses. For instance, we find that, for
 ~$M_{\eta}^{}\sim 10^{14}_{} \,\textrm{GeV}$,\,
 $\rho \sim 10^{13}_{} \,\textrm{GeV}$ and \,$y \sim \mathcal{O}(1)$,\,
 the neutrino masses can be naturally around
 ~$\mathcal{O}(0.1 \,\textrm{eV})$\,.\,
 It is clear that this new mechanism
 of the neutrino mass generation has two essential features: (i)
 it generates Dirac masses for neutrinos,
 and (ii) it retains the essence of the conventional
 seesaw\,\cite{seesaw} by making the neutrino masses tiny via the small ratio
 of the weak scale over the heavy mass scale,
 and hence may be called {\it Dirac Seesaw}.
 We further note that
 the above Dirac-type dimension-5 effective operator (\ref{dim-5D})
 does share certain essential features with the traditional
 Majorana-type dimension-5 effective operator\,\cite{Weinberg},
 \begin{eqnarray}
 \label{dim-5M}
 \dis
 {\cal L}^{(5)}_{\rm Majorana} ~=\,
 -\f{c}{\Lambda}\psi_L^{T}\epsilon\phiT\,{\cal C}\,
                \phiT^{\,T}_{}\epsilon\psi_L^{}
 \,+\, {\rm h.c.}
 \,,
 \end{eqnarray}
 where $\,\phiT =i\tau_2^{~}\phi^{\ast}\,$ has hypercharge $+\f{1}{2}$
 and $\,{\cal C}=i\gamma^2\gamma^0\,$ is the charge-conjugation
 operator.  The major difference
 of the current {\it Dirac Seesaw} from the conventional
 {\it Majorana Seesaw} \cite{seesaw}
 is that the former involves heavy doublet {\it scalar} rather than
 heavy singlet {\it Majorana fermion}. After integrating out the
 heavy fields (either the heavy messenger scalars in the Dirac Seesaw
 or the heavy right-handed neutrinos in the Majorana Seesaw),
 we see that the resulted effective
 dimension-5 operators (\ref{dim-5D}) and (\ref{dim-5M}) share an
 essential similarity regarding the generation of
 {\it small neutrino masses.}

 In the following, we demonstrate how to generate the observed baryon asymmetry
 within the above \textit{Dirac Seesaw} scenario.  It allows us to make
 use of the so-called Dirac leptogenesis mechanism\,\cite{dlrw1999}.
 As shown in \cite{dlrw1999}, since the sphalerons\,\cite{sphaleron} have no
 direct effect on the right-handed fields, a nonzero lepton asymmetry
 stored in the right-handed fields could survive above the
 electroweak phase transition and then produce the baryon
 asymmetry in the universe, although the lepton asymmetry stored in
 the left-handed fields had been destroyed by the sphalerons. For all
 the SM species, the Yukawa couplings are sufficiently strong
 to rapidly cancel the stored left- and right-handed lepton numbers.
 But the effective interactions of the right-handed Dirac neutrinos are
 exceedingly weak, and the equilibrium between left- and right-handed
 lepton numbers will not be realized until temperatures fall well
 below the weak scale. At that time the left-handed lepton number has already
 been converted to baryon number by the sphalerons.
 Hence, the leptogenesis \cite{leptogenesis} could be valid even if
 the neutrinos are not of Majorana nature \cite{ars1998}.
 For the SM with three generation
 fermions and one Higgs doublet, one finds\,\cite{dlrw1999} that the final
 baryon number, which is converted from the lepton number stored in the
 left-handed leptons, should be
 \begin{eqnarray}
 \label{bsolution}
 n_{B}^{} \,=\, n_{L}^{} \,=\, -\frac{28}{79}n_{\nu_{R}^{}}^{}\,.
 \end{eqnarray}

\begin{figure}\vspace{-4.5cm}
 \psfig{file=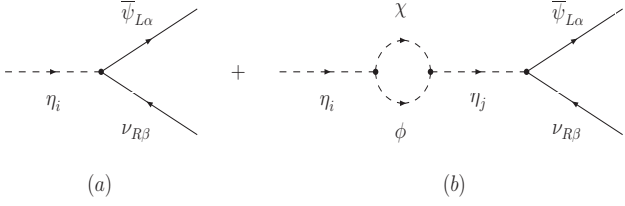, bbllx=1.8cm, bblly=4.7cm, bburx=25cm,
 bbury=30.0cm, width=12cm, height=15cm, angle=0, clip=0}
 \vspace{-6.0cm}
 \caption{\label{cpviolation} The decay of
 $\eta_{i}^{} \rightarrow \psi_{L\alpha}^{}\,\nu_{R\beta}^{c}$ at
 tree-level (a), and at one-loop order (b). The interference of these
 diagrams produces the {\tt CP}-violation needed for giving rise to
 a net asymmetry stored in the right-handed neutrinos. }
 \end{figure}

 In the Dirac leptogenesis scenario, the requirement of {\tt CP}
 violation must be fulfilled in order to generate a baryon asymmetry
 in the early universe. In the present model, the heavy messengers
 have two decay processes:
 \begin{eqnarray}
 \eta \rightarrow \left\{ \begin{array}{c}
 \psi_{L}^{}\,\nu_{R}^{c}   \\
 \\
 \phi\,\chi  \end{array} \right.,
 \quad\eta^{\dagger}_{} \rightarrow \left\{ \begin{array}{c}
 \psi_{L}^{c}\,\nu_{R}^{}   \\
 \\
 \phi^{\dagger}_{}\,\chi^{\dagger}_{}   \end{array} \right..
 \end{eqnarray}
 The channels of $\eta \rightarrow \psi_{L}^{}\,\nu_{R}^{c}$ and
 $\eta^{\dagger}_{} \rightarrow \psi_{L}^{c}\,\nu_{R}^{}$ could
 provide the expected asymmetry between the right-handed neutrinos
 and antineutrinos as long as the {\tt CP} is not conserved. For
 the purpose to produce {\tt CP}-violation, we need at least
 two such heavy messenger scalars $\eta_{j}^{}$ ($j=1,2$)
 to realize the interference between the
 tree diagram and the irreducible loop correction.
 Here we minimally introduce two heavy messengers and write
 down the relevant Lagrangian,
 \begin{eqnarray}
 \label{yt2}
 \!\mathcal{L} \supset
 \sum_{j=1}^{2}\left[-M_{\eta_{j}^{}}^{2}\eta_j^{\dagger}\etaj
 -y^{}_{j\alpha\beta}\bar{\psi}_{L\alpha}^{}\etaj\nu_{R\beta}^{}+
 \rho_{j}^{}\chi\eta^{\dagger}_{j}\phi+ \!{\rm h.c.}\right]\!.~
 \end{eqnarray}
 where the summation over the family indices $(\alpha,\,\beta)$ is
 implied.
 Fig.\,\ref{cpviolation} depicts the decay of
 $\eta_i^{}\rightarrow \psi_{L\alpha}^{}\,\nu_{R\beta}^{c}$ at tree-level and
 one-loop order.
 Then, we can define the {\tt CP} asymmetry for the
 right-handed neutrinos,
 \begin{eqnarray}
 \label{right}
 \!
 \varepsilon_{Ri}^{}\equiv
 \frac{\,\sum_{\alpha\beta}^{}\left[\Gamma(\eta^{\dagger}_{i}
 \!\rightarrow\!
 \psi_{L\alpha}^{c}\,\nu_{R\beta}^{})-\Gamma(\eta_{i}^{}\!\rightarrow\!
 \psi_{L\alpha}^{}\,\nu_{R\beta}^{c})\right]\,}{\Gamma_{i}^{}}\,,~
 \end{eqnarray}
 where \,$\Gamma(\eta_{i}^{}\rightarrow
 \psi_{L\alpha}^{}\,\nu_{R\beta}^{c})$\,
 is the decay width of
 \,$\eta_{i}^{} \rightarrow \psi_{L\alpha}^{}\,\nu_{R\beta}^{c}$\,,\,
 and so on.
 $\Gamma_{i}^{}$ denotes the total decay width of $\eta_{i}^{}$ or
$\eta_{i}^{\dagger}$,
\begin{eqnarray}
\label{ucpt}
\Gamma_{i}^{}&\equiv&\sum_{\alpha\beta}^{}\Gamma(\eta_{i}^{}\rightarrow
\psi_{L\alpha}^{}\,\nu_{R\beta}^{c})+\Gamma(\eta_{i}^{}\rightarrow\phi\,
\chi) \nonumber\\
&\equiv& \sum_{\alpha\beta}^{}\Gamma(\eta^{\dagger}_{i} \rightarrow
\psi_{L\alpha}^{c}\,\nu_{R\beta}^{})+\Gamma(\eta^{\dagger}_{i}\rightarrow
\phi^{\dagger}_{}\,\chi^{\dagger}_{})\,.
\end{eqnarray}
 Note that the second equality in (\ref{ucpt}) is guaranteed by the
 unitarity and the {\tt CPT} conservation.
 In comparison with the right-handed neutrinos, the left-handed leptons
 should have an equal but opposite {\tt CP} asymmetry. After a
 lengthy calculation, we derive the total decay width  and the
 {\tt CP} asymmetry, respectively,
 \begin{eqnarray}
 \Gamma_{i}^{}=
 \frac{1}{8\pi}\left[\textrm{Tr}(y_{i}^{\dagger}y_{i}^{})
 +\frac{\rho_{i}^{2}}{M_{\eta_{i}^{}}^{2}}\right]
 M_{\eta_{i}^{}}^{}\,,
 \end{eqnarray}
 \begin{eqnarray}
 \label{cpa}
 \varepsilon_{Ri}^{}
 =-\frac{1}{4\pi}\frac{\textrm{Im}
 [\textrm{Tr}(y_{i}^{\dagger}y_{j}^{})\rho_{i}^{}\rho_{j}^{}]}
 {\,\textrm{Tr}(y_{i}^{\dagger}y_{i}^{})M_{\eta_{i}^{}}^{2}+\rho_{i}^{2}\,}
 \frac{M_{\eta_{i}^{}}^{2}}
 {\,M_{\eta_{i}^{}}^{2}\!-\!M_{\eta_{j}^{}}^{2}\,}\,,
 \end{eqnarray}
 where $\,j\neq i\,$.

 In the case that the two heavy messengers have hierarchical masses, the
 final lepton asymmetry comes mainly from the decay of the lighter
 one. For illustration, let us focus on this hierarchical case.
 Without losing generality, we will choose $\eta_{1}^{}$ to be the
 ighter messenger and $\eta_{2}^{}$ the heavier one.
 Thus, we can simplify the {\tt CP}-asymmetry as,
\begin{eqnarray}
\label{cpa1} \varepsilon_{R1}^{}\simeq\frac{1}{4\pi}
\frac{\textrm{Im}[\textrm{Tr}(y_{1}^{\dagger}y_{2}^{})]\rho_{1}\rho_{2}^{}}
{\,\textrm{Tr}(y_{1}^{\dagger}y_{1}^{})M_{\eta_{1}^{}}^{2}+\rho_{1}^{2}\,}
\frac{M_{\eta_{1}^{}}^{2}}{M_{\eta_{2}^{}}^{2}}\,,
\end{eqnarray}
 where the cubic couplings $\rho_1$ and $\rho_2$ are properly
 defined to be real as mentioned below Eq.\,(\ref{yt}).
 The presence of the relative phases between $y_{1}^{}$ and $y_{2}^{}$
 can generate a net lepton asymmetry
 stored in the right-handed neutrinos via the
 out-of-equilibrium decay of the lighter messenger $\eta_1^{}$.
 Subsequently, as long as the transmission between the produced left-
 and right-handed asymmetries is prevented from coming into thermal
 equilibrium at least until after the electroweak phase transition,
 the right-handed asymmetry could be hidden from the sphalerons and
 then the left-handed lepton number gets converted into the baryon asymmetry.

 It is reliable to determine the final baryon asymmetry by exactly solving
 the full Boltzmann equations, which
 is beyond the scope of the present demonstration and
 will be presented elsewhere.
 For convenience, we will adopt the conventional approximate
 expression for the asymmetry of the right-handed neutrinos\,\cite{kt1980}:
 \begin{eqnarray}
 \label{lsolution}
 Y_{\nu_{R}^{}}^{}\equiv\frac{n_{\nu_{R}^{}}^{}}{s}\simeq \left\{\!
 \begin{array}{lc}
 \varepsilon_{R1}^{}/g_{\ast}^{} \,,~
 &\textrm{for}~ K \ll
 1, \\
 \\
 0.3\,\varepsilon_{R1}^{}/[g_{\ast}^{}K(\ln K)^{0.6}_{}] \,,~
 &\textrm{for}~ K \gg 1,
 \end{array} \right.
 \end{eqnarray}
 where $s$ is the entropy density and ~$g_{\ast}^{} =
 \mathcal{O}(100)$~ is the number of relativistic degrees of freedom
 in the early universe. The parameter $K$ stands for the ratio of the
 total decay width of $\eta_{1}^{}$ to the expansion rate of the
 universe:
 \begin{eqnarray}
 \label{parameter}
 K\equiv\left.\frac{\Gamma_{1}^{}}{2H(T)}\right|_{T=M_{\eta_{1}^{}}}^{}
 =\left( \frac{16\pi^{3}_{}g_{\ast}^{}}{45} \right)^{\!\!-\f{1}{2}}_{}
 \frac{M_{\rm Pl}\Gamma_{1}^{}}{M_{\eta_{1}^{}}^{2}}\,,
 \end{eqnarray}
 where
 \,$H(T)= (8\pi^{3}_{}g_{\ast}^{}/90)^{1/2}_{}T^{2}_{}/M_{\rm Pl}^{}$\,
 is Hubble constant and
 $\,M_{\rm Pl}^{}\simeq 1.22 \times 10^{19}_{}$\,GeV\,
 is Planck mass.

 It is clear from Eq.\,(\ref{cpa1}) that we must have at least two
 heavy messengers for the $\nu_{R}^{}-\nu_{R}^{c}$ asymmetry to be
 nonzero, hence the neutrino mass matrix described in Eq.\,(\ref{diracmass})
 should be expanded as,
 \begin{eqnarray}
 \label{diracmass2}
 M_{\nu}^{}= vu\left[
 \frac{y_{1}^{}\rho_{1}^{}}{M_{\eta_{1}^{}}^{2}}+
 \frac{y_{2}^{}\rho_{2}^{}}{M_{\eta_{2}^{}}^{2}}\right]\,.
 \end{eqnarray}
 Under the limit of $M_{\eta_{1}^{}}\ll M_{\eta_{2}^{}}$ and for
 $\rho_{i}^{}\lesssim M_{\eta_{i}^{}}$, we can
 neglect the contribution from $\eta_{2}^{}$ in the above equation.
 Thus, we further derive the neutrino masses as,
 \begin{eqnarray}
 \label{diracmass3}
 \sum_{j=1}^{3}m_{j}^{2}
 &=&\textrm{Tr}(M_{\nu}^{\dagger}M_{\nu}^{})
 \simeq\textrm{Tr}(y_{1}^{\dagger}y_{1}^{})
 \frac{\rho_{1}^{2}v^{2}_{}u^{2}_{}}{M_{\eta_{1}^{}}^{4}}\nonumber\\
 &=&\frac{(4\pi)^{5}_{}g_{\ast}^{}}{45}B_{f}^{}B_{s}^{}K^{2}_{}
 \frac{v^{2}_{}u^{2}_{}}{M_{\rm Pl}^{2}}\,,
 \end{eqnarray}
 where \,$m_{j}^{}$ ($j=1,2,3$) are the neutrinos mass-eigenvalues.
 $B_{f}^{}$ and $B_{s}^{}$
 are the branching ratios of
 $\eta_{1}^{}$ decaying to the fermions and the scalars,
 respectively,
 \begin{eqnarray}
 \label{bf}
 \hspace*{-7mm}
 B_{f}^{}&\equiv&
 \frac{\,\sum_{\alpha\beta}^{}\Gamma(\eta_{1}^{}\rightarrow
 \psi_{L\alpha}^{}\,\nu_{R\beta}^{c})\,}{\Gamma_{1}^{}}
 =\frac{\textrm{Tr}(y_{1}^{\dagger}y_{1}^{})M_{\eta_{1}^{}}^{}}{8\pi\Gamma_{1}^{}}\,,\\
 \label{bs}
 \hspace*{-7mm}
 B_{s}^{}&\equiv&
 \frac{\Gamma(\eta_{1}^{}\rightarrow
 \phi\,\chi)}{\Gamma_{1}^{}}
 =\frac{\rho_{1}^{2}/M_{\eta_{1}^{}}^{}}{8\pi\Gamma_{1}^{}}\,.
 \end{eqnarray}
 We note that $B_{f}^{}$ and $B_{s}^{}$ obey the relationship,
 \begin{eqnarray}
 B_{f}^{}+B_{s}^{} = 1\,,\quad \Longrightarrow \quad
 B_{f}^{}B_{s}^{}\leqq \frac{1}{4}\,.
 \end{eqnarray}
 It is convenient to express $K$ as a function of the neutrino
 masses,
 \begin{eqnarray}
 K=\left[\frac{(4\pi)^{5}_{}g_{\ast}^{}}{45}\right]^{\!-\frac{1}{2}}_{}
 \left(B_{f}^{}B_{s}^{}\right)^{\!-\frac{1}{2}}
 \frac{\,\mB M_{\rm Pl}\,}{vu}\,,
 \end{eqnarray}
 where the quadratic mean $\mB$ of the neutrino masses is defined
 by ~$\mB^{2}_{}\equiv\dis\sum_{j=1}^{3}m_{j}^{2}$\,.\,

 For illustration below, let us choose,
\begin{eqnarray}
\label{assumption}
r\equiv\frac{M_{\eta_{1}^{}}^{}}{M_{\eta_{2}^{}}^{}}
=\frac{\rho_{1}^{}}{\rho_{2}^{}}\,,\quad y_{2}^{}=
y_{1}^{}e^{i\delta}_{}\,,
\end{eqnarray}
 where $\delta$ denotes the relative phase between $y_{1}^{}$
 and $y_{2}^{}$. So, we can derive the {\tt CP}-asymmetry from
 Eq.\,(\ref{cpa1}),
 \begin{eqnarray}
 \label{cpa2}
 \varepsilon_{R1}^{}
 &=&\frac{1}{4\pi}\frac{\textrm{Tr}(y_{1}^{\dagger}y_{1}^{})\rho_{1}^{2}}
 {\textrm{Tr}(y_{1}^{\dagger}y_{1}^{})M_{\eta_{1}^{}}^{2}+\rho_{1}^{2}}
 r\sin\delta\,
 \nonumber\\[2mm]
 &=&\left[\frac{(4\pi)^{3}_{}g_{\ast}^{}}{45}\right]^{\!\frac{1}{2}}_{}
 B_{f}^{}B_{s}^{}K\frac{M_{\eta_{1}^{}}}{M_{\rm Pl}^{}}r\sin\delta
 \nonumber\\[2mm]
 &=&\dis\frac{(B_{f}^{}B_{s}^{})^{\frac{1}{2}}_{}}{4\pi}
 \frac{\,M_{\eta_{1}^{}}\mB\,}{vu}r\sin\delta\, .
 \end{eqnarray}
 For the inputs of \,$B_{f}^{}B_{s}^{}= 1/4$\,,\,
 \,$\sin\delta = - 1$,\, \,$r= 0.1$,\,
 \,$u=1\,\textrm{TeV}$,
 \,\,$M_{\eta_{1}^{}}= 1.15\times 10^{12}\,\textrm{GeV}$\, and\,
 $\mB = 0.1 \,\textrm{eV}$\,,\,
 we derive the sample predictions,
 \,$\varepsilon_{R1}^{} \simeq -2.6\times
 10^{-6}_{}$\, and \,$K\simeq 16$\,.\,
 In consequence, applying Eqs.\,(\ref{bsolution}) and
 (\ref{lsolution}),
 we arrive at, ~$n_{B}^{}/s\simeq 8.7\times 10^{-11}$\,.\,
 From this we can deduce,
 $~n_{B}^{}/n_{\gamma}^{}\simeq 7.04 n_B^{}/s
   \simeq 6.1\times 10^{-10}\,$,\,
 which agrees well to the
 observation \cite{spergel2006},
 ~$n_{B}^{}/n_{\gamma}^{}=(6.0965\pm 0.2055)\times 10^{-10}$\,,\,
 where $\,n_{\gamma}^{}\,$ is the current photon number density.

 In conclusion, we have presented a new {\it Dirac Seesaw} scenario to
 simultaneously explain the origin of neutrino masses and the
 generation of baryon asymmetry in the universe.
 This is achieved by extending the
 visible content of the SM with a hidden sector which has new
 $U(1)_{X}^{}$ gauge symmetry and is composed of three
 right-handed singlet neutrinos and one singlet Higgs.
 The messages between these two sectors
 are mediated by the heavy messenger scalar doublets. The heavy
 messengers can help to highly suppress the neutrino masses and also decay
 efficiently to provide the baryon asymmetry. Further phenomenological
 consequences from the hidden sector (including the new gauge boson $Z'$ from
 the broken $U(1)_X$ \cite{U1})
 are interesting and will be studied elsewhere.

 \vspace*{3mm}
 \noindent

\appendix
\section{}
 The general scalar potential in our model is
 \begin{equation}
 \begin{array}{l}
 V(\phi,\chi,\eta_{i}^{}) 
 \\[2mm]
 = -\mu^{2}_{}\phi^{\dagger}_{}\phi
 +\lambda(\phi^{\dagger}_{}\phi)^{2}_{}-
 \sigma^{2}_{}\chi^{\dagger}_{}\chi+\kappa(\chi^{\dagger}_{}\chi)^{2}_{}
 +\xi\phi^{\dagger}_{}\phi\chi^{\dagger}_{}\chi
 \\[2.5mm]
 \hspace*{5mm}
 +\dis\sum_{ijk\ell}^{}\{M_{\eta_{i}^{}}^{2}\eta_{i}^{\dagger}\eta_{i}^{}+
 \delta_{ijk\ell}\eta_{i}^{\dagger}\eta_{j}^{}\eta_{k}^{\dagger}\eta_{\ell}^{}
 \\[2.5mm]
 \hspace*{5mm}
 +\,\alpha_{ij}^{}\phi^{\dagger}_{}\phi\eta_{i}^{\dagger}\eta_{j}^{}
 +\beta_{ij}^{}\phi^{\dagger}_{}\eta_{i}^{}\eta_{j}^{\dagger}\phi+
 \gamma_{ij}^{}\chi^{\dagger}_{}\chi\eta_{i}^{\dagger}\eta_{j}^{}
 \\[2.5mm]
 \hspace*{5mm}
 -\,[\,\rho_{i}^{}\chi\eta_{i}^{\dagger}\phi \,+\,{\rm h.c.}\,]\}\,,
 \end{array}
 \end{equation}
 where the indices $(i,j,k,\ell )=1,2,\cdots$, denote the number of the messenger
 doublets. Without losing generality, the tree-level mass matrix for
 the scalar doublets $\eta_{i}$ have been chosen to be diagonal and
 real. We then analyze the potential as a function of the scalar VEVs,
 \,$v\equiv\langle\phi \rangle$\,,\,
 \,$u\equiv\langle\chi\rangle$\, and
 \,$w_{i}^{}\equiv\langle\eta_{i}^{} \rangle$\,,
 \begin{eqnarray}
 && \hspace*{-12mm} V(v,u,w_{i}^{})
 \nonumber\\
 \hspace*{-6mm}
 &=& -\mu^{2}_{}v^{2}_{}+\lambda
 v^{4}_{}-\sigma^{2}_{}u^{2}_{}+\kappa u^{4}_{}+\xi
 v^{2}_{}u^{2}_{}
 \nonumber\\
 \hspace*{-6mm}
 &&+\sum_{ijk\ell}^{}\{M_{\eta_{i}^{}}^{2}w_{i}^{2}+
   \delta_{ijkl}w_{i}^{}w_{j}^{}w_{k}^{}w_{\ell}^{}
   \\
 \hspace*{-6mm}
 &&+(\alpha_{ij}^{}+\beta_{ij}^{})v^{2}_{}w_{i}^{}w_{j}^{}+
 \gamma_{ij}u^{2}_{}w_{i}^{}w_{j}^{}-2\rho_{i}^{}vuw_{i}^{}\}\,.
 \nonumber
 \end{eqnarray}
 Taking the extreme conditions, ~$0=\partial V/\partial v=\partial
 V/\partial u=\partial V/\partial w_{i}^{}$\,,\, we obtain
 \begin{eqnarray}
 \label{vevphi}
 \hspace*{-4mm}
 0&=&-2\mu^{2}_{}v+4\lambda v^{3}_{}+2\xi
 vu^{2}_{}\nonumber\\
 \hspace*{-4mm}
 &&+\sum_{ij}^{}\left[
 2(\alpha_{ij}^{}+\beta_{ij}^{})vw_{i}^{}w_{j}^{}-2\rho_{i}^{}uw_{i}^{}\right],
 \\
 \label{vevsigma}
 \hspace*{-4mm}
 0&=&-2\sigma^{2}_{}u+4\kappa u^{3}_{}+2\xi
 v^{2}_{}u\nonumber\\
 \hspace*{-4mm}
 && +\sum_{ij}^{}\left[2\gamma_{ij}uw_{i}^{}w_{j}^{}-2\rho_{i}^{}vw_{i}^{}\right],\\
 \label{veveta}
 \hspace*{-4mm}
 0&=&\sum_{jk\ell}^{}\left[
 2M_{\eta_{i}^{}}^{2}w_{i}^{}+4\delta_{ijk\ell}w_{j}^{}w_{k}^{}w_{\ell}^{}\right.
 \nonumber\\
 \hspace*{-4mm}
 && \left. +2(\alpha_{ij}^{}+\beta_{ij}^{})v^{2}_{}w_{j}^{}+
 2\gamma_{ij}u^{2}_{}w_{j}^{}-2\rho_{i}^{}vu\right].
 \end{eqnarray}
 As before, we will choose ~$M_{\eta_{i}^{}}\gg
 v,u$~ and ~$\rho_{i}^{} \lesssim M_{\eta_{i}^{}}$\,,\, and thus we
 deduce $w_{i}^{}$ \cite{ms1998} from solving Eq.\,(\ref{veveta}),
 \begin{eqnarray}
 w_{i}^{}~\simeq~ \frac{\,\rho_{i}^{}vu\,}{M_{\eta_{i}^{}}^{2}} ~\ll~ v,u\,.
 \end{eqnarray}
 Subsequently, we have
 \begin{eqnarray}
 v^{2}_{}&=&
 \frac{\,2\kappa\mu^{2}_{}-c\sigma^{2}_{}\,}
      {4\lambda\kappa-c^{2}_{}}\,,
 \\[2mm]
 u^{2}_{}&=&
 \frac{\,2\lambda\sigma^{2}_{}-c\mu^{2}_{}\,}
      {4\kappa\lambda-c^{2}_{}}\,,
 \end{eqnarray}
 where the coefficient ~$c\equiv
 \xi -\dis\sum_{i}^{}\rho_{i}^{2}/M_{\eta_{i}^{}}^{2}$\,.\,
 In order to ensure $v$ and $u$ to be nonzero, we can
 choose the parameters \,$\mu^{2}_{}$,
 $\sigma^{2}_{}$, $\lambda$, $\kappa$\, and \,$c$\, to
 satisfy the following conditions,
 \begin{eqnarray}
 \hspace*{-4mm}
 \left\{\begin{array}{l} c \,\neq\, 0 \,,
 \\[2mm]
 4\kappa\lambda-c^{2}_{} \,>\, 0 \,,
 \\[2mm]
 \dis
 \f{c}{2\kappa}\,<\, \f{\mu^{2}}{\sigma^{2}} \,<\, \f{2\lambda}{c} \,,
 \end{array}\right.
 ~~{\rm or,}~~~
 c~=~0\,.
 \end{eqnarray}

 \vspace*{20mm}


\begin{thebibliography}{99}

 \bibitem{Frank}
 A different and inspiring approach to the hidden sector was recently
 suggested by B. Patt and F. Wilczek, {\tt hep-ph/0605188}.


 \bibitem{seesaw}
 P. Minkowski, Phys. Lett. B {\bf 67}, 421 (1977); T. Yanagida, in
 {\it Proceedings of the Workshop on Unified Theory and the Baryon
 Number of the Universe}, edited by O. Sawada and A. Sugamoto (KEK,
 Tsukuba, 1979), p. 95; M. Gell-Mann, P. Ramond, and R. Slansky, in
 {\it Supergravity}, edited by F. van Nieuwenhuizen and D. Freedman
 (North Holland, Amsterdam, 1979), p. 315; S.L. Glashow, in {\it
 Quarks and Leptons}, edited by M. L$\rm\acute{e}$vy {\it et al.}
 (Plenum, New York, 1980), p. 707; R.N. Mohapatra and G.
 Senjanovi$\rm\acute{c}$, Phys. Rev. Lett. {\bf 44}, 912 (1980).

 \bibitem{Weinberg}
 S. Weinberg, Phys. Rev. Lett. {\bf 43}, 1566 (1979).

\bibitem{dlrw1999}
K. Dick, M. Lindner, M. Ratz, and D. Wright, Phys. Rev. Lett. {\bf
84}, 4039 (2000); H. Murayama and A. Pierce, Phys. Rev. Lett. {\bf
89}, 271601 (2002); B. Thomas and M. Toharia, Phys. Rev. D {\bf 73},
063512 (2006); D.G. Cerdeno, A. Dedes, and T.E.J. Underwood, JHEP
{\bf 0609}, 067 (2006); B. Thomas and M. Toharia, hep-ph/0607285.


\bibitem{sphaleron}
V.A. Kuzmin, V.A. Rubakov and M.E. Shaposhnikov, Phys. Lett. B {\bf
155}, 36 (1985); R.N. Mohapatra and X. Zhang, Phys. Rev. D {\bf 45},
2699 (1992).

\bibitem{leptogenesis}
M. Fukugita and T. Yanagida, Phys. Lett. B {\bf 174}, 45 (1986); P.
Langacker, R.D. Peccei, and T. Yanagida, Mod. Phys. Lett. A {\bf 1},
541 (1986); M.A. Luty, Phys. Rev. D {\bf 45}, 455 (1992); R.N.
Mohapatra and X. Zhang, Phys. Rev. D {\bf 46}, 5331 (1992). For a
recent review, W. Buchmuller, R.D. Peccei, and T. Yanagida, Ann.
Rev. Nucl. Part. Sci. {\bf 55}, 311 (2005).


\bibitem{ars1998}
Dirac neutrinos are also suitable in the mechanism of baryogenesis
via neutrino oscillations, E.Kh. Akhmedov, V.A. Rubakov, and A.Yu.
Smirnov, Phys. Rev. Lett. {\bf 81}, 1359 (1998).


\bibitem{kt1980}
E.W. Kolb and M.S. Turner, \textit{The Early Universe},
Addison-Wesley, Reading, MA, 1990; H.B. Nielsen and Y. Takanishi,
Phys. Lett. B {\bf 507}, 241 (2001).




 \bibitem{spergel2006}
 D.N. Spergel {\it et al.}, {\tt astro-ph/0603449}.

 \bibitem{U1}
 For a recent review, T.G. Rizzo, {\tt hep-ph/0610104}.

  \bibitem{ms1998}
 Our mechanism of generating small VEVs for heavy scalar doublets is
 similar to the type-II seesaw model where heavy scalar triplets get
 small VEVs for the mass generation of Majorana neutrinos,
 E. Ma and U. Sarkar, Phys. Rev. Lett. {\bf 80}, 5716 (1998).


\end{thebibliography}
\end{document}